# Field Programmable Gate Array based Front-End Data Acquisition Module for the COSMICi Astroparticle Telescope System

Darryl W. McGowan, Jr., David R. Grosby, Michael P. Frank, Sachin Junnarkar, Ray H. O'Neal, Jr.

*Abstract*—**We describe an FPGA based Front-End Data Acquisition Module (FEDAM) for implementing Time-over-Threshold (ToT) Time-to-Digital conversion (TDC) of pulses obtained from the COSMICi astroparticle telescope detector system photomultiplier tubes. The telescope consists of a minimum of three scintillation detectors configured to detect particle airshowers likely initiated by Ultra High Energy Cosmic Rays (UHECR). The relative time delay of detection events between the detectors is used to estimate the angle of incidence of the shower. The FEDAM provides time-over-threshold measurements with a resolution of 2 ns. This allows determination of shower direction to an error of 0.035 (cos $\theta$)$^{-1}$ radians where $\theta$ is the angle between the baseline axis through a pair of detectors and the plane representing the shower front.**

## I. INTRODUCTION

Time over Threshold (ToT) techniques for determination of high energy particle detector output pulse characteristics is increasingly being applied or under consideration for use in both space and ground based high energy particle and particle astrophysics experiments [1], [2], [3]. In contrast to the more traditional method of analog to digital converter (ADC) based systems for pulse acquisition and digitization [4], ToT approaches are well suited for use with time to digital converters (TDCs) implemented on reconfigurable computing hardware such as Field Programmable Gate Arrays (FPGAs). Whereas in the recent past, ToT/TDC systems might have required application specific integrated circuits (ASICs) for implementation, the increasing availability of FPGAs avoids the cost and complication of the ASIC design process [5]. The re-programmability of FPGAs also provides flexibility in terms of real-time filtering and analysis of the detector pulses according to evolving science priorities of a particular experiment. Additionally, using multiple voltage thresholds in the ToT system provides two data points for pulse shape reconstruction at every voltage threshold level crossed. In the COSMICi Front-End Data Acquisition Module (FEDAM) described herein, a ToT-TDC system implemented using an FPGA is used to estimate the orientation of the front of extensive air showers (EAS) initiated by interactions between high energy astroparticles and the atmosphere of the Earth and to estimate the charge content of the photomultiplier tube output pulses (to be discussed in detail in a subsequent publication). In section II, we describe the FEDAM architecture and hardware design. In section III, we describe the method by which the TDC is implemented on the FPGA. Finally, in section IV, we present time resolution performance data for the FEDAM.

## II. FEDAM BOARD DESIGN

### A. FEDAM Layout

Figure 1.1 shows the FEDAM electronics board developed through a collaboration between the AstroParticle and Cosmic Radiation Detector Research and Development Laboratory (APCR-DRDL) of the Physics Department at Florida A&M University (FAMU) and Brookhaven National Laboratory. Layout and fabrication of the board was performed in the Instrumentation Division of Brookhaven National Laboratory. The development of the electronics was pursued to facilitate participation of FAMU in the MARIACHI experiment [6].

Voltage pulses from the scintillation detector PMTs are input to the FEDAM through SMA ports located on the left side of the board. The input signal is compared to voltage level thresholds provided by programmable Digital to Analog Converters (DACs) located on the board.

Comparators in low-voltage differential signaling (LVDS) input pins of the FPGA determine whether the input pulse voltage on each input channel is greater than each DAC level.

This work was supported in part by the U.S. National Science Foundation (NSF) under the CREST program, award number 0630370. We also gratefully acknowledge donations of components and equipment received from Altera Corporation's University Program.

D. W. McGowan, Jr. and D. R. Grosby are students in the Department of Electrical & Computer Engineering at the FAMU-FSU College of Engineering, Tallahassee, FL 32310 USA (219-381-4003; email: darryl1.mcgowan@famu.edu, preferred emails).

M. P. Frank is a Research Associate with the AstroParticle & Cosmic Radiation Detector Research & Development Laboratory and an adjunct instructor in the Department of Electrical & Computer Engineering at the FAMU-FSU College of Engineering, Tallahassee, FL 32310 USA (michael.frank@famu.edu).

S. Junnarkar is currently with Toshiba Medical Research Institute (junnarkar@gmail.com).

R. H. O'Neal, Jr. is director of the AstroParticle & Cosmic Radiation Detector Research & Development Laboratory (APCR-DRDL) and Associate Professor of Physics, Department of Physics, Florida A&M University, Tallahassee, FL 32304 USA (ray.oneal@famu.edu)

By setting the DACs to specific threshold levels and recording the times at which each input crosses them, pulses can be represented by a series of bits that represent the shape and time of arrival of input pulses. After the data has been acquired, it is then transferred from the board via a UART serial connection for further processing.

### B. FPGA Architecture

Fig 1.2 provides a system level diagram of the FPGA architecture. The COSMICi FEDAM FPGA is a model Stratix II donated to FAMU through the Altera University Program (AUP). The FPGA Architecture consists of three main top level modules, The High-Speed Time Counter (HSTC), a multi-channel pulse-form input-capture datapath, and the Nios II Microcontroller System. The HSTC counts cycles of a 500 MHz on-chip clock derived from the 50 MHz system clock using a phase-locked loop (PLL). The Pulse Capture Datapaths (PCDPs) use the HSTC to timestamp the instants at which the rising and falling edges of the input pulse cross each voltage threshold. The absolute time of each pulse arrival event is determined by counting the number of cycles of a reference high-precision oscillator occurring between GPS 1 pulse per second (1 PPS) clock signals. A soft-core Nios II microcontroller system, developed using the Altera SOPC (system-on-a-programmable chip) Builder tool, is used to run the application firmware.

### C. Time-over-Threshold Implementation

The Time-over-Threshold technique is used to represent a pulse by the amount of time that the pulse magnitude exceeds some defined threshold level or levels. The applications of ToT techniques for pulse processing are described in many references [1], [2].

Custom peripheral device interfaces were designed to program the DACs to allow the threshold levels to be controlled in software. Currently, the threshold levels are logarithmically spaced over the expected range of pulse heights; however, sufficient flexibility exists to space threshold levels arbitrarily according to linear, polynomial, logarithmic or other distributions appropriate for adequate pulse shape reconstruction.

### III. FPGA BASED TDC

The original plan for implementation of the TDC in the FPGA was based on a ring oscillator design developed by one of the authors (Junnarkar, S.) [7]. Difficulty in reproducibly setting ring oscillator frequencies required pursuing a simpler design described hereafter. Fig. 3.1 and Fig. 3.2 aid in discussion of the TDC conversion.

As shown in the aforementioned figures, the counter value is latched when the threshold levels are crossed from below and above the threshold. By calculating the difference in the latched counter states for each threshold, the time that the input exceeded each threshold can be computed. Combining the pulse width information obtained above along with the voltage magnitude provided by the threshold crossings, the pulse shape can be reconstructed.

### IV. TIME RESOLUTION PERFORMANCE

#### A. Overall Performance

Presently, the maximum sustained rate at which the system can transmit pulse-shape data to the server is measured to be approximately 175 pulses per second, limited by the baud rate of the serial communication link, although this could be improved upon by compressing the data encoding.

The time resolution of the system is defined as the minimum time increment, as measured, between two subsequent digitized time values [8]. For example, a 500 MHz clock yields a time resolution of 2 ns, resulting in a standard deviation for the error of pulse width measurements of approximately 1.15 ns. With respect to a HSTC value, the actual time of threshold crossing of the rising and falling edge of an arbitrary input pulse are randomly distributed within a 2-ns-wide-window according to a uniform probability distribution. The 1.15 ns standard deviation was obtained by analyzing the triangular distribution of the input signal pulse widths for their difference (assuming independence of the rise and fall times). Fig. 4.1 shows the probability distributions used in the analysis. Each voltage threshold crossing time is registered within a uniformly distributed range of 2 ns, providing the triangular distribution from which the error was calculated.

Fig. 4.2 and 4.3 show test results of the FEDAM when supplied an input pulse from an external waveform generator. Multiplying the digitized time values by the time resolution and plotting against the threshold values, Fig. 4.3 is constructed.

#### B. Determination of Shower Front Orientation

The time difference between detection events occurring between scintillators in a given shower provides a simple estimate of the shower front orientation. See Fig. 4.4 for an illustration of the basic approach in a simplified two-detector scenario.

The figure depicts the detection time delay,

$$t = \frac{r \sin \theta}{c}, \qquad (1)$$

between two detectors A and B separated by a distance $r$, due to a shower event in which the angle between the shower front and the baseline axis between detectors is $\theta$. $c$ is the speed of light, which is appropriate for the highly relativistic particles in high-energy EAS showers. The error in angle is determined from taking the differential of both side of the time delay equation as follows

$$\delta t = \frac{r(\cos\theta)\delta\theta}{c}. \qquad (2)$$

Therefore,

$$\delta\theta = \frac{c\delta t}{r\cos\theta}. \qquad (3)$$

For a detector separation distance of 10 meters and a pulse width standard deviation of 1.15 ns, we obtain for the error in angle, 0.035/cos $\theta$ radians.

## ACKNOWLEDGMENTS

We would like to acknowledge contributions from Paul O'Connor of The Instrumentation Division of Brookhaven National Laboratory for facilitating the fabrication of the FEDAM. Also, we would like to thank Helio Takai for helpful discussions and continuing collaboration.

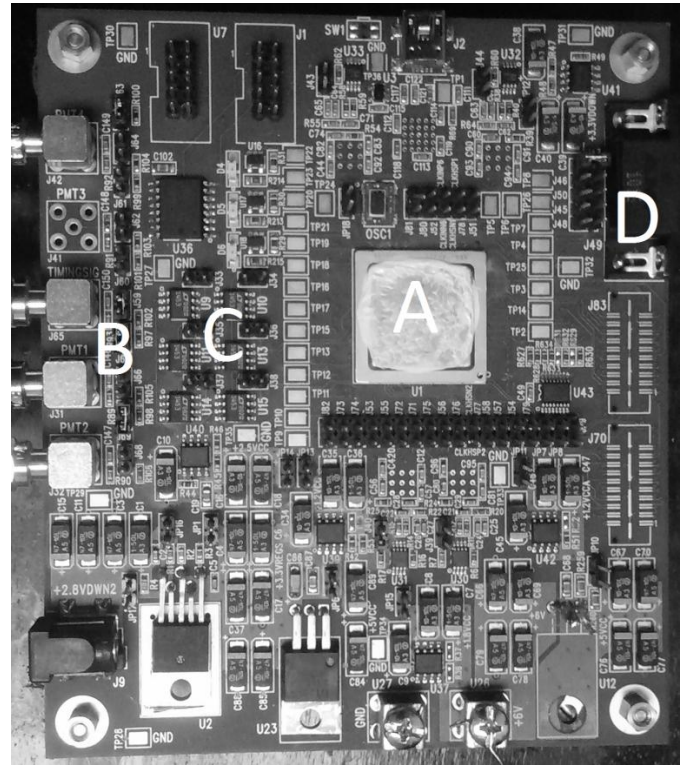

Fig. 1.1(a): Front-End Data Acquisition Module. A: Stratix II FPGA (thermal paste applied for cooling system). B: SMA port location for PMT inputs. C: Digital-to-Analog Converters. D: DE9 connector

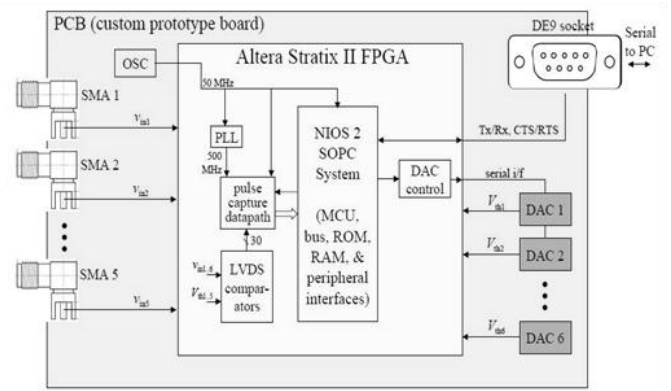

Fig. 1.1(b): System level diagram of FEDAM. This diagram shows the flow of data though each module of the FEDAM.

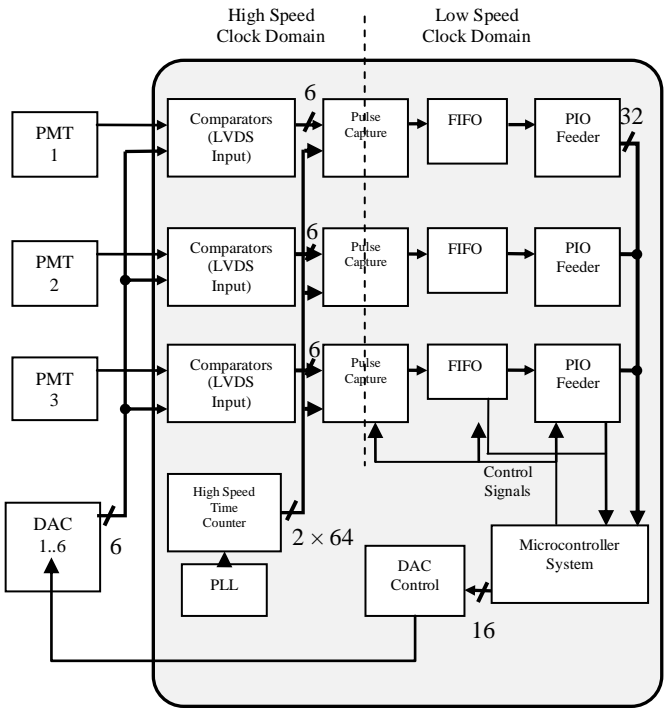

Fig. 1.2: Top Level FPGA Architecture Schematic. As shown, two clocks govern the timing of the system, both of which are used in the FPGA—the 50 MHz FPGA system clock and the designed PLL clock used for generating the high speed counter for high time resolution.

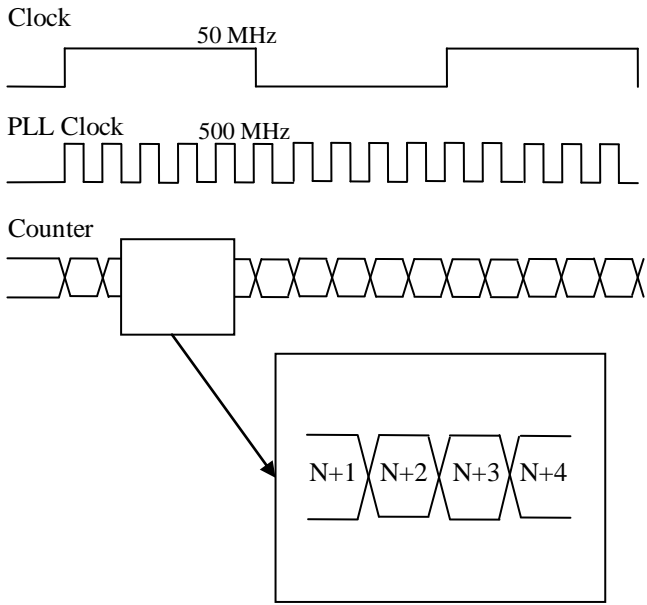

Fig. 3.1: Timing diagram representing system clock and high-speed Phase Locked Loop (PLL) clock. The inset depicts high-speed counter sequence from system start.

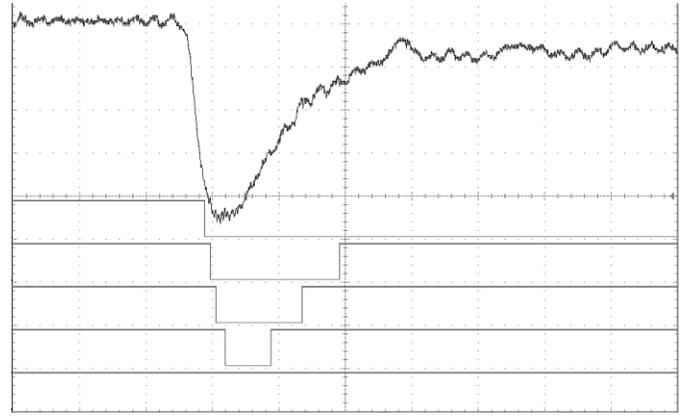

Fig. 3.2: Diagram depicting a typical analog input pulse along with its resulting digital representation. The threshold voltages compared against the input signal can be selected by dynamic system calibration. The comparator outputs shown below the input pulse are logic high when the input voltage is greater than that of the corresponding threshold and logic low otherwise.

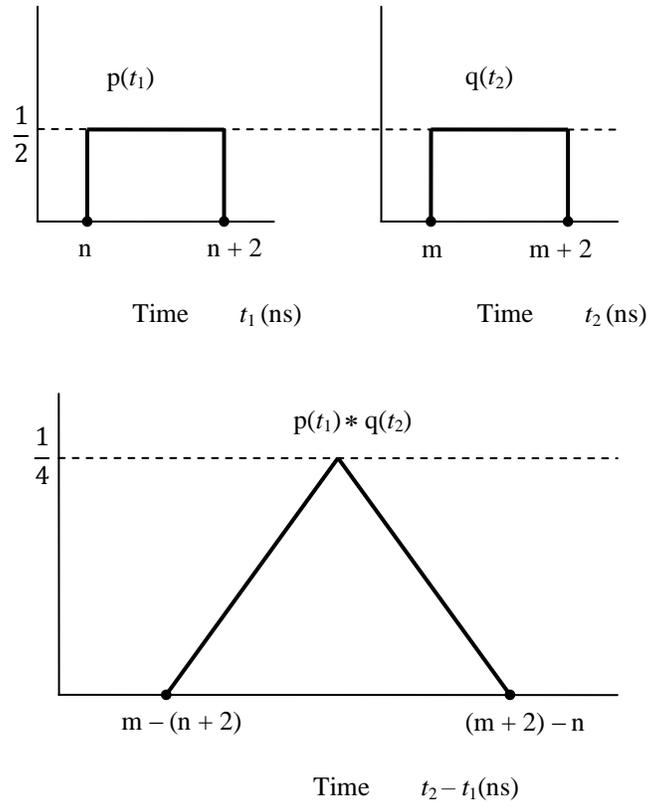

Fig. 4.1: Diagram that depicts the functions p(t) and q(t), the uniform probability distributions of the rising and falling edge threshold crossing times of input pulses, respectively. After convolution of the two distributions,

$$\int_{-\infty}^{\infty} q(t-\tau)p(\tau)d\tau$$

, the triangular probability distribution q(t) ✳ p(t) is yielded. Using the time resolution of 2 ns, the standard deviation of this triangular probability distribution is 1.15 ns.

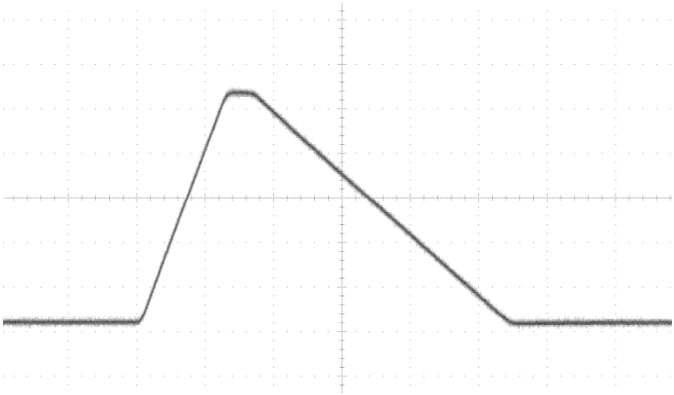

Fig. 4.2: Pulse generated by external waveform generator with a nominal pulse height of +2.6V, 20/60 ns leading/trailing edge transition times, and 58 ns pulse width. (In this trace, the horizontal scale is 20 ns/division and vertical is 0.5V/div.)

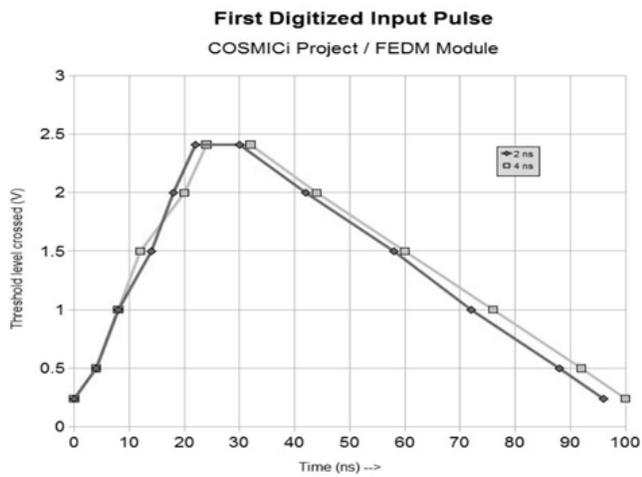

Fig. 4.3: Plot of results of data from test in Figure 4.3. Blue line was reconstructed with a time resolution of 4 ns. Red line with resolution of 2 ns. The six threshold levels used in this test were 0.24, 0.50, 1.00, 1.50, 2.00, and 2.41V respectively.

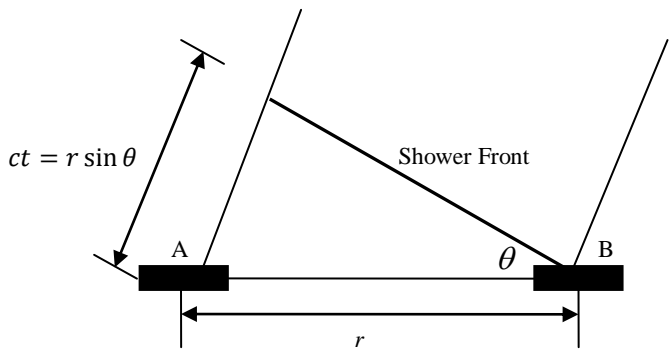

Fig. 4.4: Diagram depicting a two-detector configuration labeled with parameters used to determine the orientation of the shower front.